\documentclass[a4paper,11pt]{article}
\pdfoutput=1 

\usepackage{jinstpub} 

\title{BabyMOSS stitched sensors: results of characterisation tests for ALICE ITS3 upgrade}

\author[1]{Alessandro Sturniolo\note{Speaker}}
\collaboration[c]{on behalf of the ALICE collaboration}

\affiliation{INFN Sezione di Catania,\\Via Santa Sofia 64 - 95123 Catania, Italy}
\affiliation{MIFT Department, University of Messina,\\Viale Ferdinando Stagno d'Alcontres 31 - 98166 Messina, Italy}

\emailAdd{a.sturniolo@cern.ch}

\abstract{During the Long~Shutdown~3 (LS3, scheduled 2026-2030), the innermost 3~layers (Inner Barrel, or IB) of the present ALICE ITS2 will be replaced with large-area, flexible, stitched CMOS 65~nm sensors, arranged in 2~half-barrels of 3 half-layers each, in the framework of the ITS3 upgrade project. For the first time in a High Energy Physics experiment, such large-scale sensors will be bent into truly half-cylindrical shaped half-layers, requiring little mechanical support. This will also help lower the material budget: a reduction down to an average of 0.09\%~X$_{0}$ per layer is expected, benefitting ITS tracking and vertexing capabilities especially at low momenta. 

In the wafer yield and stitching assessment stage for ITS3 R\&D, test devices from the Engineering Run~1 (ER1) submission were developed, including the MOnolithic Stiched Sensors (MOSS) and smaller variants of them (babyMOSS). In particular, babyMOSS is a single Repeated Sensor Unit (RSU) of a MOSS device: the chip is $\sim$14$\times$30~mm$^{2}$ in size, and consists of 8~digitally read~out pixel~matrices (regions) arranged in 2~rows, i.e. half-units (HUs). 

BabyMOSS chip characterisation tests have been performed in laboratory and test beam environments. Laboratory tests include systematic functional scans to study the behaviour of front-end electronics over a range of different settings, whereas test beam measurements, under high-energy charged particle beams, were used to investigate the detection efficiency and spatial resolution. 

In this contribution, we will present the babyMOSS chip characterisation campaign, with a focus on recent test beam results. So far, babyMOSS test beam results have been consistent with full MOSS, and confirmed that babyMOSS devices meet the ITS3 requirements: a detection efficiency $>$ 99\%, fake hit rate $<$ 10$^{-6}$ hits per pixel and event, and a spatial resolution $<$ 6 $\upmu$m.}


\proceeding{26$^{\text{th}}$ International Workshop on Radiation Imaging Detectors \\
  6--10 July 2025 \\
  Bratislava}

\notoc

\begin{document}
\maketitle
\flushbottom

\section{Introduction}
\label{sec:intro}
\begin{figure}[h!]
	
	\centering
	\includegraphics[width=0.45\linewidth]{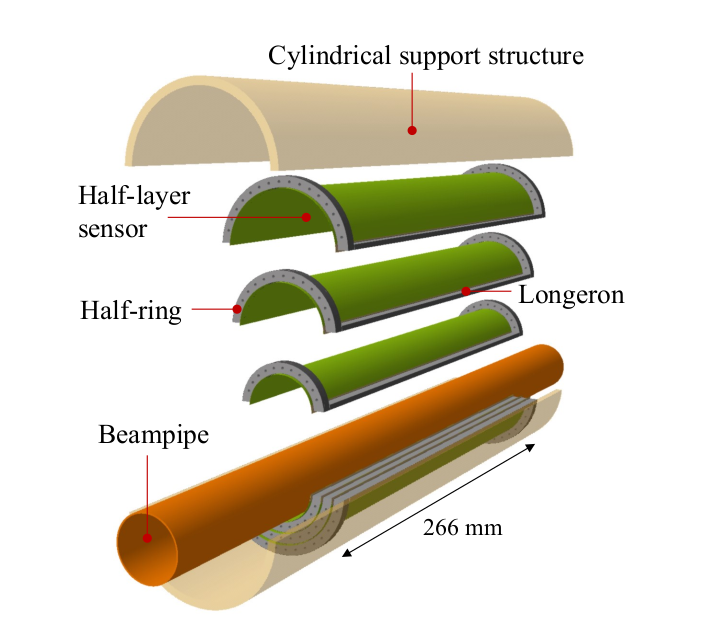}
	\hfill
    \includegraphics[width=0.42\linewidth]{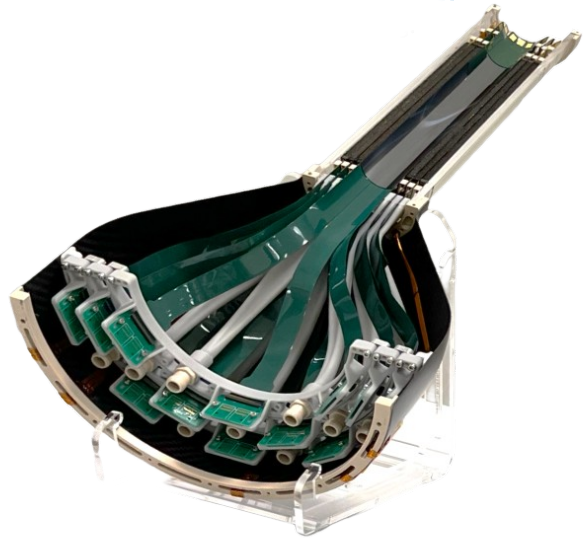}
	\caption{Layout of the new ITS3 (left) and a partially integrated model of ITS3, built both with commercial and final-grade components (Engineering Model 3, or EM3)~\cite{its3-tdr}.}
	\label{fig:its3_layout_section}

\end{figure}

During the Long Shutdown 3 (LS3), the ALICE Inner Tracking System (ITS) will be upgraded to a new version, called the ITS3. For the first time in a High Energy Physics experiment, the ITS3 will feature truly cylindrical tracking layers, that will replace the 3~innermost layers (also known as Inner Barrel, or IB) of the state-of-the-art ITS2. Each layer will consist of 2~large-scale, stitched sensors (half-layers), produced with a CMOS 65~nm technology and bent into a half-cylindrical shape. The structure will be held in place by a carbon foam support with longerons, as shown in figure~\ref{fig:its3_layout_section}, where a mockup of the upgraded ITS3 is also presented. 

With the upcoming ITS3, the material budget of the ALICE ITS will drop from the current 0.36\%~X$_{0}$ per layer to an average of just 0.09\%~X$_{0}$, which will contribute to enhancing the ITS3 tracking capabilities at low transverse momenta ($p_{T}$). The installation of an air cooling system, set to replace the current water cooling, will also contribute to this decisive material budget reduction.

Several tracks of preliminary tests have been run and are still ongoing to date, in order to maximise the information gain on the full ITS3 design, including (but not limited to) sensor bending, aeromechanical and cooling tests, and small-scale and stitched prototype characterisation~\cite{its3-loi,its3-tdr}.

\section{Testing stiched sensors for ITS3: Engineering Run 1 and babyMOSS}
\label{sec:er1_babyMOSS}
\begin{figure}[h!]
	
	\begin{minipage}{0.48\linewidth}
		\includegraphics[width=0.98\linewidth]{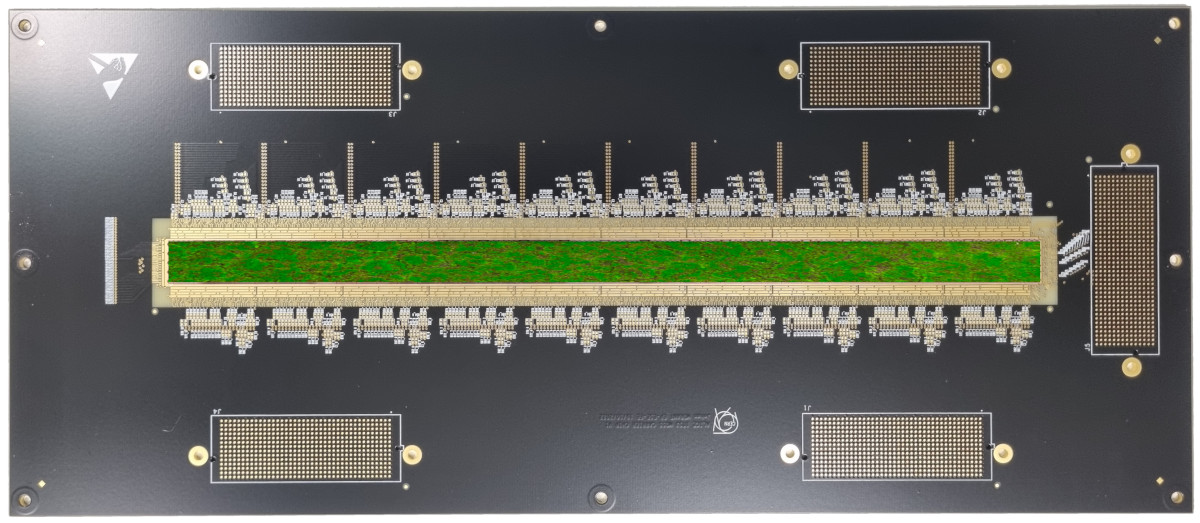}
		\includegraphics[width=0.98\linewidth]{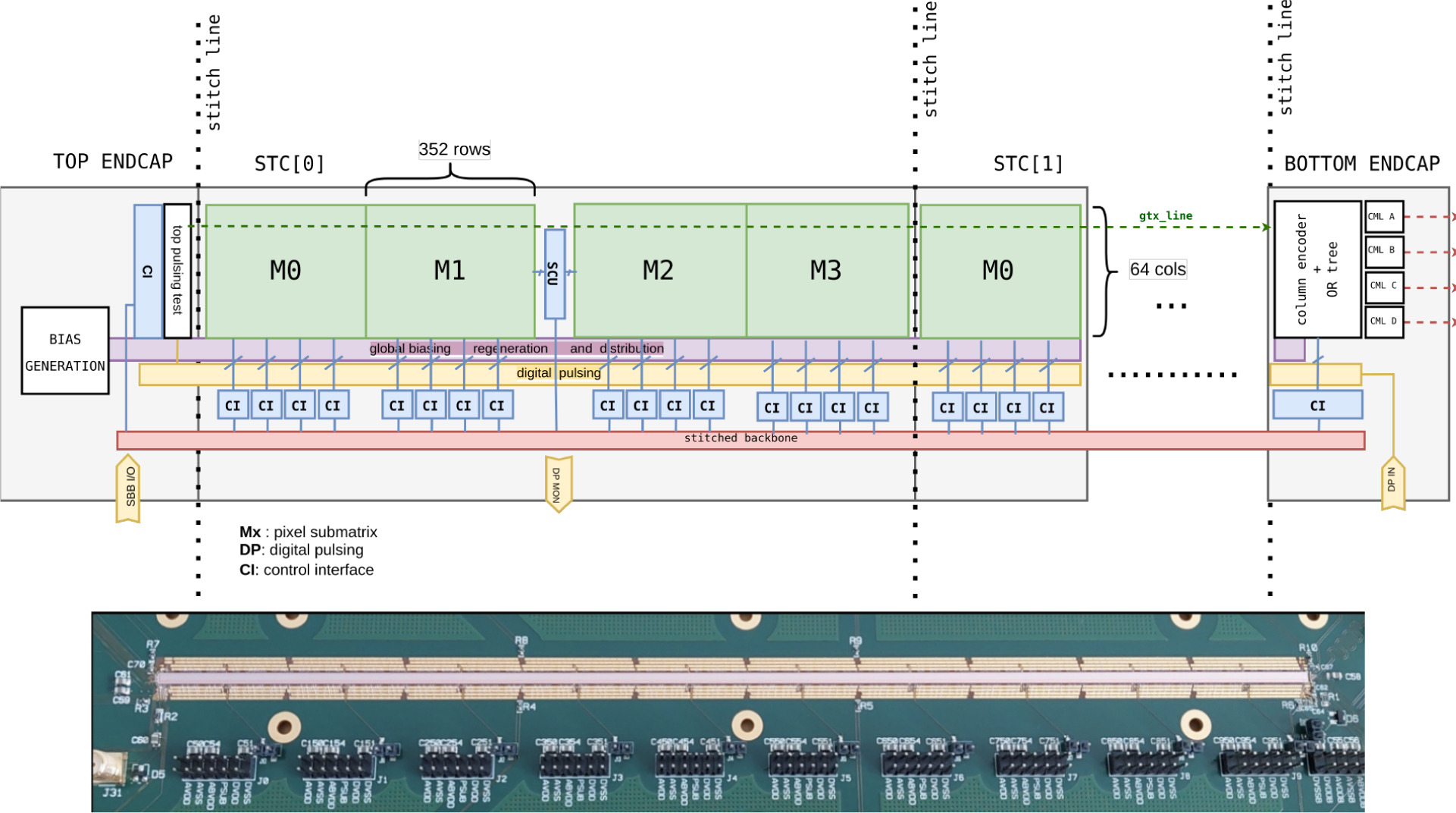}
	\end{minipage}
    \hfill
    \begin{minipage}{0.48\linewidth}
    	\includegraphics[height=0.33\textheight]{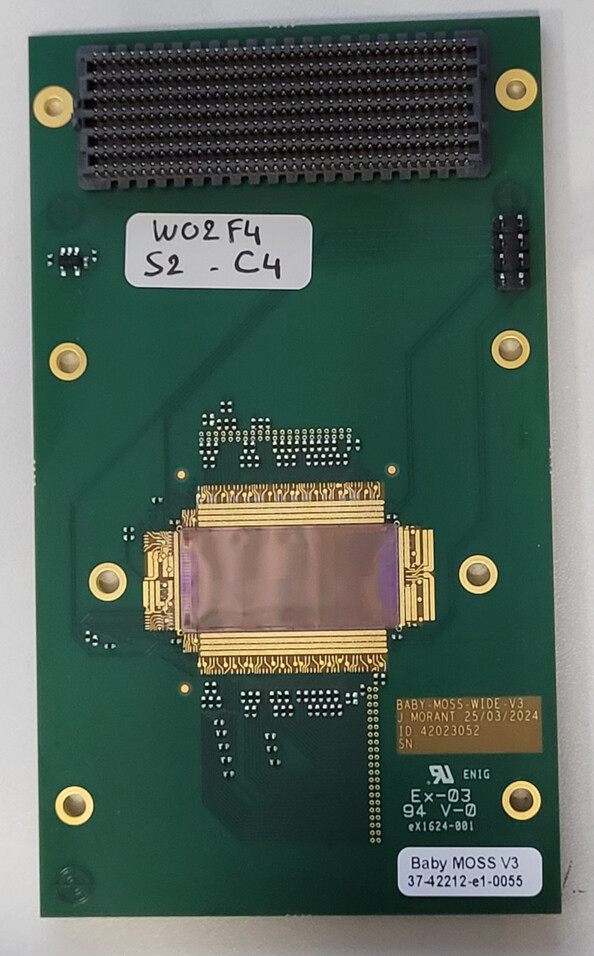}
    \end{minipage}
    \caption{ER1 devices: picture of a full MOSS (top left), functional diagram and picture of a MOST (bottom left), and picture of a babyMOSS chip (right).}
    \label{fig:er1_devices}
    
\end{figure}

The stitching technology to be used for the new ITS3 sensors allows the manufacturing of sensors larger than the design reticle, at wafer production level~\cite{its3-loi,its3-tdr}. In order to qualify the stitching technology, stitched sensor prototype performances have been extensively studied, both in laboratory and in test beam environments. The test devices developed for this purpose were submitted in the Engineering Run~1
(ER1), in 2022; in particular, the first ER1 devices were the Monolithic Stitched Sensors (MOSSs) and Monolithic Stitched sensors with Timing (MOSTs). Both devices are large-scale chips ($\sim$26~cm long), and take 2~different approaches in terms of powering and readout system. The MOSS (figure~\ref{fig:er1_devices}, top left) has a synchronous readout and consists of 10~Repeated Sensor Units (RSUs), that can be independently powered and read out~\cite{terlizzi_pixel2024}; on the other hand, the MOST (figure~\ref{fig:er1_devices}, bottom left) has a higher pixel density and is operated with a global, conservatively designed asynchronous readout, that includes a finely granular switching system to power off defective regions~\cite{selina_pixel2024}.

Another category of ER1 devices, the babyMOSS, replicates the MOSS architecture in a smaller chip of approximately 14$\times$30~mm$^{2}$ area. The babyMOSS (figure~\ref{fig:er1_devices}, right) mainly consists of a single MOSS~RSU, and 2~end-caps on the short edges for powering and readout. 
The smaller size of babyMOSS, in comparison to full MOSS chips, makes them more convenient to irradiate and study in a test beam environment. Furthermore, the babyMOSS requires a much more compact test system, named DAQ-Raiser, requiring only a DAQ and a Raiser board, that acts as an interface between the babyMOSS carrier and the DAQ board.

Each babyMOSS chip includes 8~pixel matrices, also called regions, arranged in 2~rows, or half-units. 
Half-units are labeled as top and bottom, and regions within each half-unit are numbered from 0 to 3. 
Region dimensions and pixel pitch vary depending on the half-unit: top regions are 256$\times$256 matrices of 22.5~$\upmu$m pixels, whereas bottom regions are 320$\times$320 matrices, with a smaller pixel pitch of 18~$\upmu$m. This pixel size difference within the same chip allows to compare and optimise the device performances as a function of the pixel pitch.

\section{September~2024 babyMOSS test beam campaign: setup and goals}
\label{sec:sep24_beamtest}
Several test beam campaigns have been run on MOSS and babyMOSS devices since August~2023, aiming to investigate the detection efficiency and spatial resolution of these stitched chips at varying front-end settings. The goal was to find an operational margin with high detection efficiency ($>99\%$) and low fake-hit rate (hereafter abbreviated to FHR, $<10^{-6}$~hits per pixel and event). 

One of these beam tests was run in September~2024, at the CERN Proton Synchrotron (PS) facility, to study the performances of babyMOSS chips from different wafer production processes (splits) and at varying irradiation levels, and assess their radiation hardness. 
For this purpose, the Devices Under Test (DUTs) were targeted with 7 and 10~GeV/c charged pion beams\footnote{Initially, the beam momentum was set to 7~GeV/c due to a misconfiguration error. Later on, the momentum was set to its expected value of 10~GeV/c, but data taken with the 7~GeV/c beam is still available.}. 

\begin{figure}[h!]
	
	\centering
	\begin{minipage}{0.42\linewidth}
		\includegraphics[width=0.98\linewidth, angle=270]{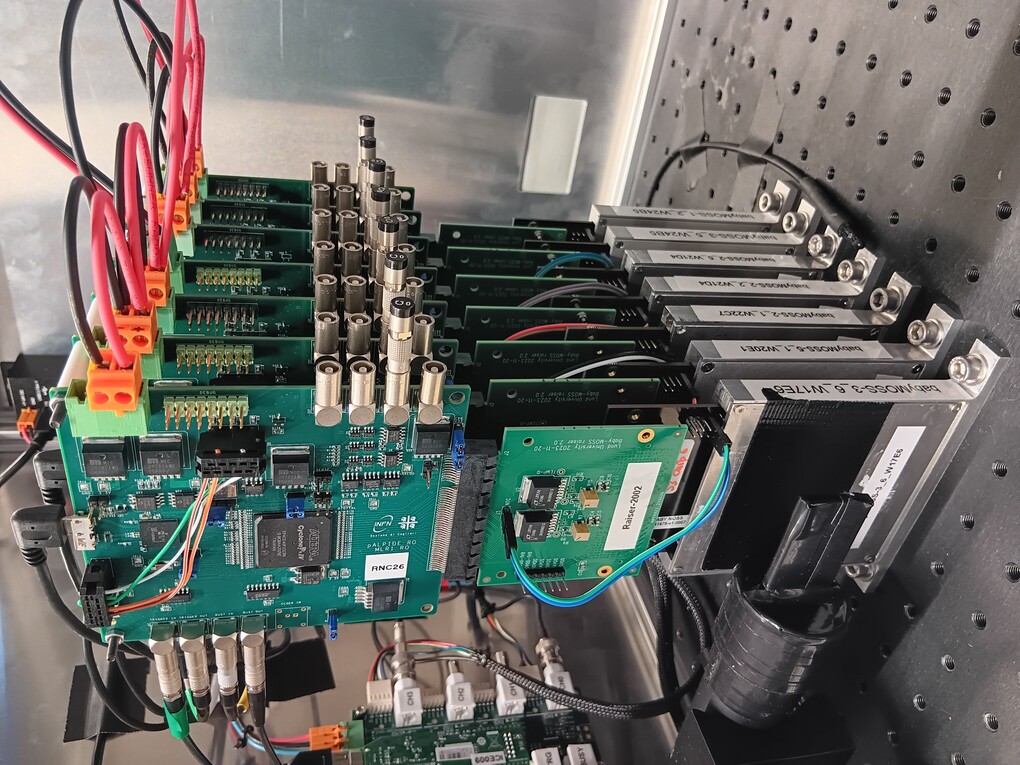}
	\end{minipage}
	\begin{minipage}{0.56\linewidth}
		\includegraphics[width=0.98\linewidth, trim={0.7cm 0 0.3cm 0}, clip]{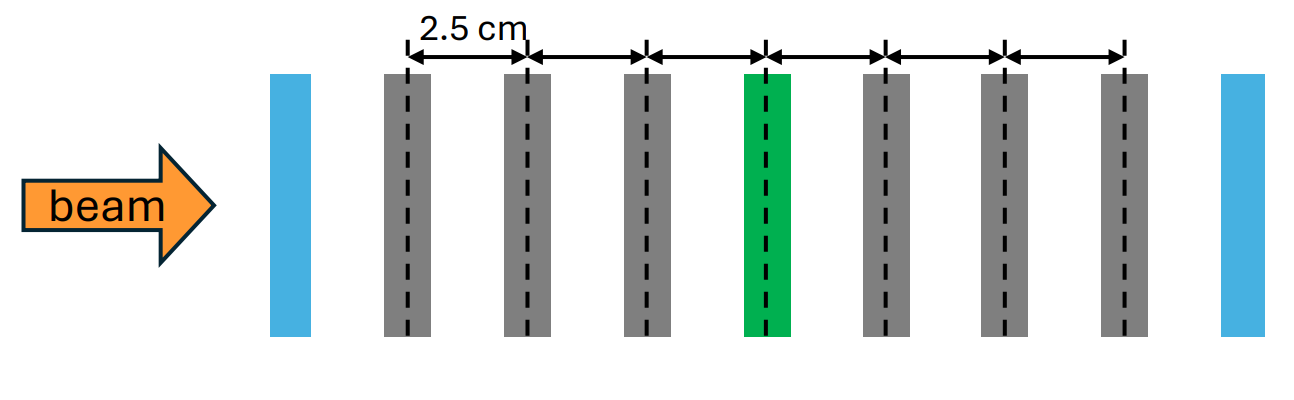}
		\vfill
		\includegraphics[width=0.98\linewidth, trim={0.7cm 0 0.3cm 0}, clip]{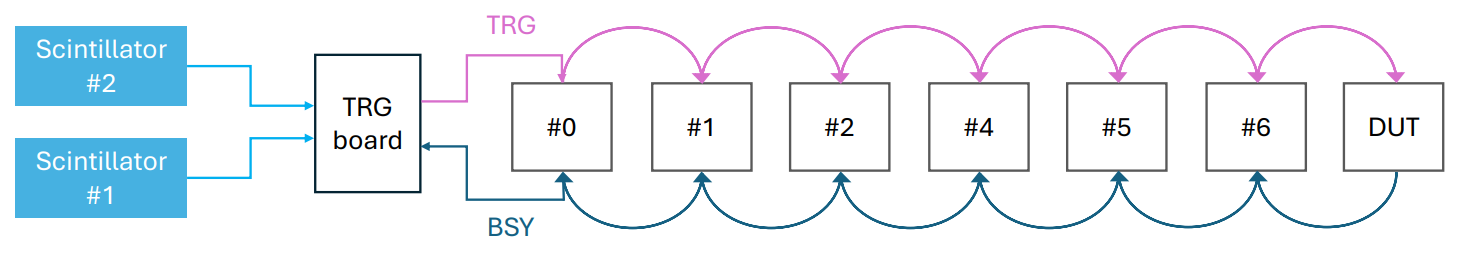}
	\end{minipage}
	\caption{September 2024 test beam telescope: picture of the setup (left), scheme and trigger chain (right). Trigger scintillators are highlighted in blue, tracking planes in gray, and the DUT in green.}
	\label{fig:sep24_TB_telescope}
\end{figure}

The beam test was run in parasitic mode; a picture of the September~2024 test beam telescope is shown in figure~\ref{fig:sep24_TB_telescope}, on the left panel. On the right side, a diagram of the setup geometry (top) and trigger chain (bottom) are presented. The telescope consisted of 6~babyMOSS chips as tracking planes (highlighted in gray in the figure), a DUT (green) positioned in between, and 2~scintillators (blue) providing a trigger to the system. All the babyMOSS planes (tracking and DUT) were mounted with the DAQ-Raiser system. 

Each trigger would be started by a coincidence of signals from the scintillators: the trigger would propagate through all tracking planes in sequence, eventually reaching the DUT. After that, a busy signal would be sent from the DUT, backwards through the other tracking planes and to the trigger board, preventing new triggers until it turns off~\cite{panasenko_bttb13,pantouvakis_bttb13}. 

\begin{table}[h!]
	\centering
	\caption{September~2024 beam test: babyMOSS DUT list, including their split and irradiation level (if irradiated).}
	\begin{tabular}{|l|l|l|}
		\hline
		{\bf Chip ID} & {\bf Split} & {\bf Irradiation} \\
		\hline
		babyMOSS-2$\_$1$\_$W22C7 & 2 & None \\
		\hline
		babyMOSS-2$\_$2$\_$W02F4 & 1 & $10^{13}$~1~MeV~n$_{\mathrm{eq}}$cm$^{-2}$~NIEL \\
		\hline
		babyMOSS-3$\_$3$\_$W02F4 & 1 & $10^{13}$~1~MeV~n$_{\mathrm{eq}}$cm$^{-2}$~NIEL \\
		\hline
	\end{tabular}
	\label{tab:sep24_TB_DUTs}
\end{table}

The DUTs investigated in this test beam campaign are listed in table~\ref{tab:sep24_TB_DUTs}. The first chip to be tested, babyMOSS-2$\_$1$\_$W22C7, was from split~2, and was the only non-irradiated device among the ones studied in September~2024. The other 2~DUTs, from split~1, had been pre-irradiated with neutrons, at a NIEL level of $10^{13}$~1~MeV~n$_{\mathrm{eq}}$cm$^{-2}$, which was a first estimate of the future ITS3 irradiation levels (today estimated around $4\cdot10^{12}$~1~MeV~n$_{\mathrm{eq}}$cm$^{-2}$). 
The goal of our test beam campaign, in particular, was to study the detection efficiency, FHR, spatial resolution and average cluster size of these DUTs, and confirm whether they can withstand the expected ITS3 radiation load.

\section{Data analysis}
\label{sec:data_analysis}
Test beam data is analysed with Corryvreckan, a lightweight, open source reconstruction and analysis framework developed for a variety of different test beam setups and DUTs~\cite{dannheim_corryvreckan}. 
Corryvreckan framework includes a variety of modules for different track reconstruction, clustering and analysis tasks. Each analysis takes as input a list of the required modules with their settings, provided in a configuration file, and the test beam telescope geometry. The geometry description must not only contain the tracking plane position and orientation, but also the plane roles: namely, a telescope plane can be flagged as a DUT, or as the reference plane, of which one (and only one) is required for telescope alignment. 

In the September~2024 analysis, the Corryvreckan workflow consisted of a sequence of noisy pixel masking, pre-alignment, alignment, and DUT analysis tasks~\cite{pantouvakis_bttb13,dannheim_corryvreckan}. On tracking planes, pixels were flagged as noisy by Corryvreckan if their hits exceeded 1000~times the average. On the DUT, noisy pixel masks were found by off-beam FHR measurements~\cite{pantouvakis_bttb13}. 

Pre-alignment and alignment tasks are typically run to adjust the plane position and orientation\footnote{The setup is first adjusted manually during the test beam preparation, but further re-adjustments down to a few $\upmu$m scale are necessary for a precise tracking, at a length scale comparable with the babyMOSS pixel pitch.}. In our case, the default right-handed cartesian coordinate system was used to describe plane positions, with the $z$-axis aligned with the beam. Pre-alignment phase involved a shift of the tracking planes along the $x$ and $y$ directions (perpendicular to the beam axis) to minimise the correlations, that in Corryvreckan are defined as the differences between cluster $x$ and $y$~positions on any given tracking plane and the reference one~\cite{dannheim_corryvreckan}. Alignment is first run without the DUT, and then including it. The full alignment procedure directly involves track reconstruction: the algorithm used by Corryvreckan starts building tracks from all possible combinations of hits in the first and last tracking plane, associating with each track any hit or cluster that falls within a certain distance. 
In our analysis, the association windows were set to 200 and 300~$\upmu$m for the alignment without and with DUT, respectively. Only tracks with at least 7 associated hits, or one per plane, were accepted as good. Planes are shifted and rotated in order to best match hits and reconstructed tracks. 

Finally, DUT analysis involved the evaluation of detection efficiency, spatial resolution and average cluster size. Hits were associated to a signal cluster within a window of 100~$\upmu$m, and DUT detection efficiency was evaluated as the ratio of hits with associated clusters over the total number of hits. Spatial resolution was instead estimated as the standard deviation of the distances between clusters and track intercepts on the DUT plane. 

\section{Results and discussion}
\label{sec:results_n_discussion}
We will now review a selection of the results from this test beam campaign. In particular, the following discussion will focus on a comparison between the non-irradiated split~2 babyMOSS-2$\_$1$\_$W22C7 DUT and one of the two irradiated DUTs, the split~1 babyMOSS-2$\_$2$\_$W02F4. The other irradiated chip has also shown similar performances to the babyMOSS-2$\_$2$\_$W02F4. 
This comparison will be limited to bottom regions, since the top regions of split~1 and 2 have different designs, but the bottom regions share the same design in both splits\footnote{In babyMOSS pixels, a low-dose n-type blanket is implanted below the collection diode, with gaps opening at the pixel edges. For top regions, the gaps are 2.5~$\upmu$m wide in split~1 and 5.0~$\upmu$m wide in split~2. For bottom regions, they are 2.5~$\upmu$m wide in both splits.}.

\begin{figure}[h!]
	
	\centering
	\includegraphics[width=0.8\linewidth]{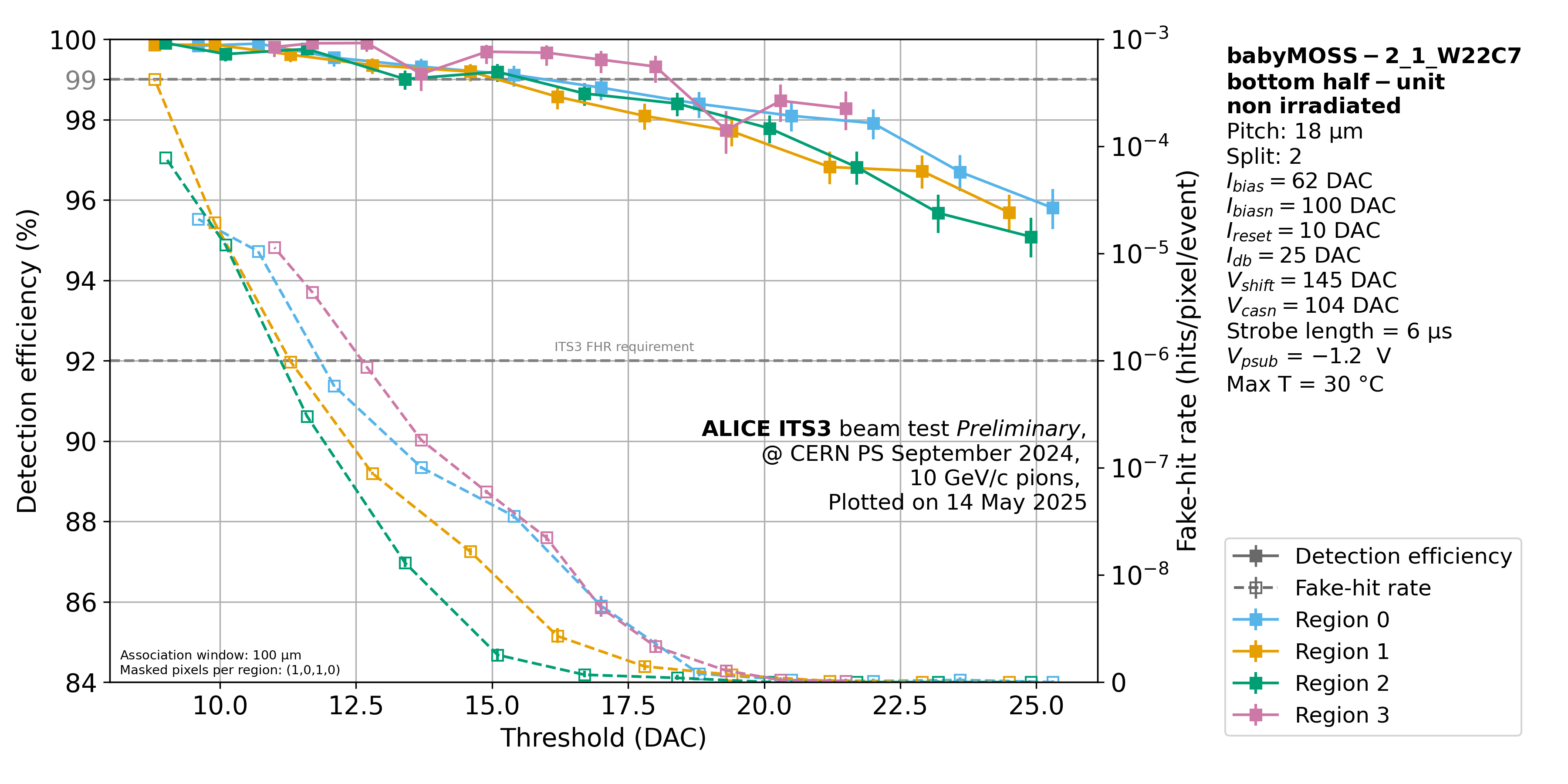}
	\includegraphics[width=0.8\linewidth]{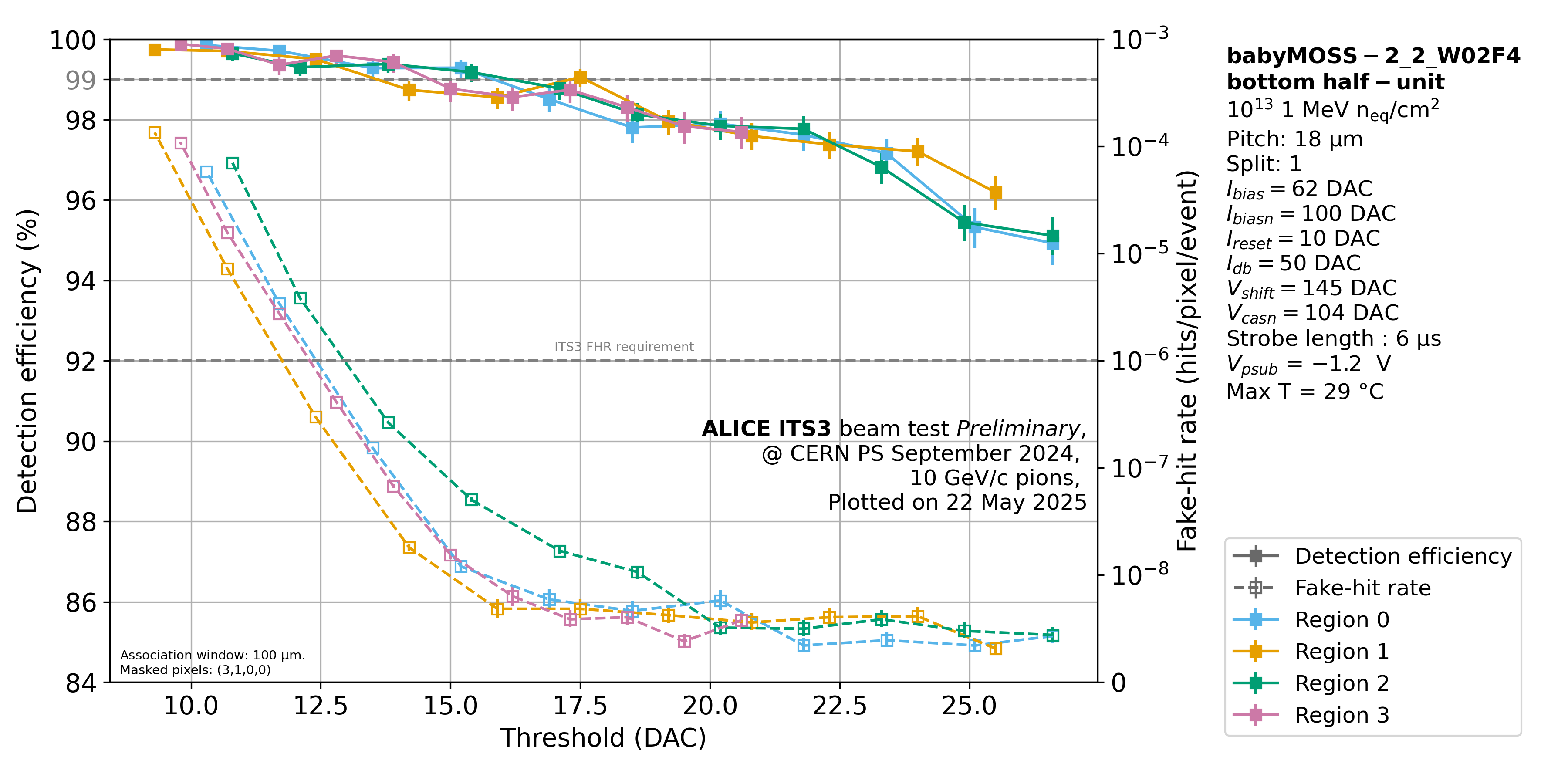}
	
	\caption{Detection efficiency (solid lines) and FHR (dashed lines) of all regions of the non-irradiated split~2 babyMOSS-2$\_$1$\_$W22C7 (top) and the irradiated split~1 babyMOSS-2$\_$2$\_$W02F4 (bottom), bottom half-unit.}
	\label{fig:deteff_noirr_vs_irr}
	
\end{figure}

In figure~\ref{fig:deteff_noirr_vs_irr}, the detection efficiency and FHR curves of both DUTs are presented as functions of the detection threshold (expressed in Digital-to-Analog Converter units, or DAC), with each colour identifying a different region. 
An operational margin with detection efficiency $> 99\%$ and FHR $< 10^{-6}$ hits per pixel and event can be found in all regions of both chips, although it is slightly narrower in the irradiated one. 
Detection efficiency appears barely affected by this irradiation level ($10^{13}$~1~MeV~n$_{\mathrm{eq}}$cm$^{-2}$~NIEL), as the efficiency curves of both irradiated and non-irradiated DUT are comparable. At threshold levels below 20~DAC, FHR does not seem to change either. However, an effect of irradiation is visible at thresholds above 20~DAC, where the FHR of the irradiated device reaches approximately $10^{-8}$ hits per pixel and event; in contrast, no FHR~noise was detected in the non-irradiated babyMOSS. This increase in FHR is due to higher leakage currents, that become more relevant as a noise source after chip irradiation.  

\begin{figure}[h!]
	
	\centering
	\includegraphics[width=0.8\linewidth]{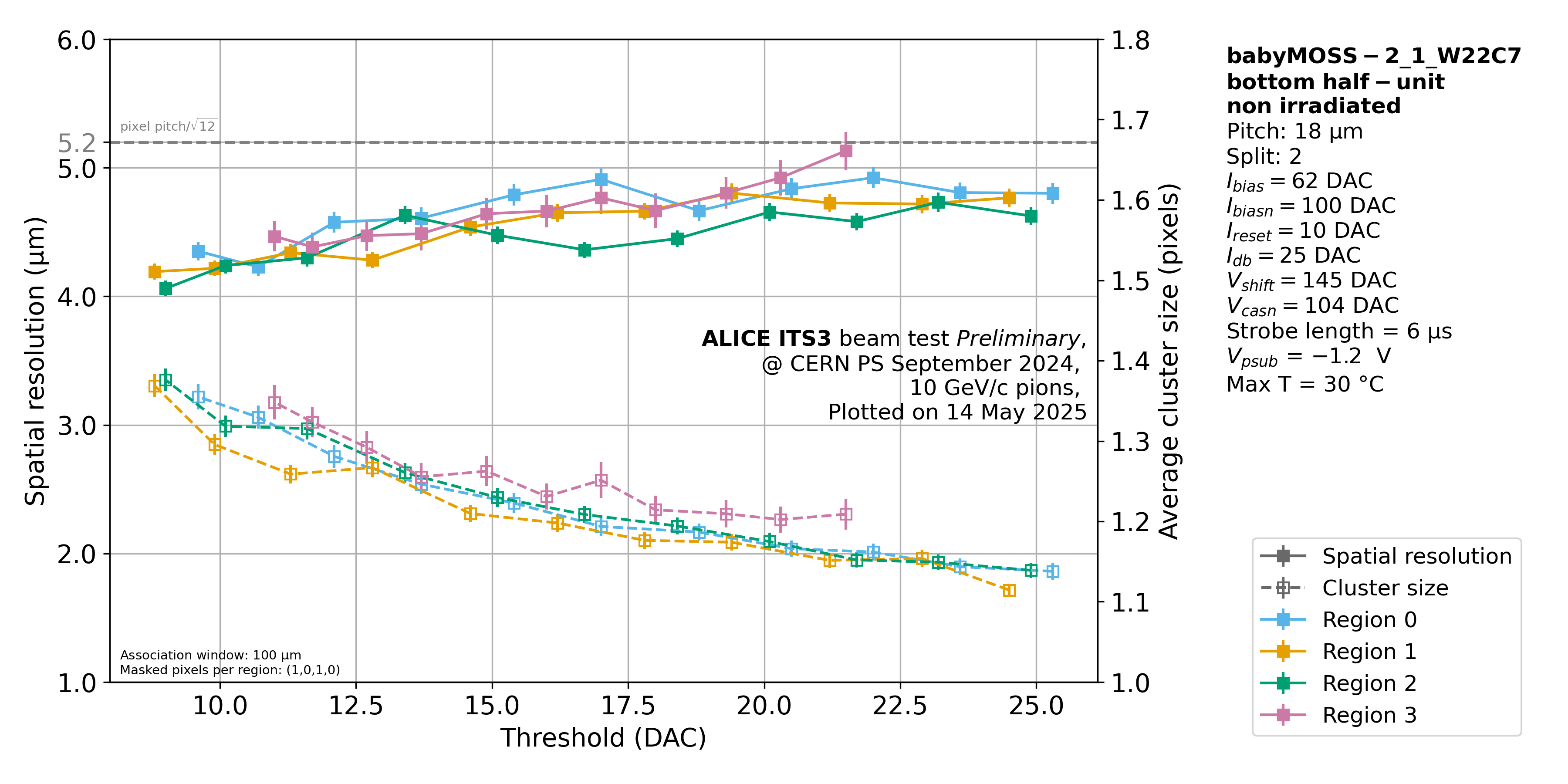}
	\includegraphics[width=0.8\linewidth]{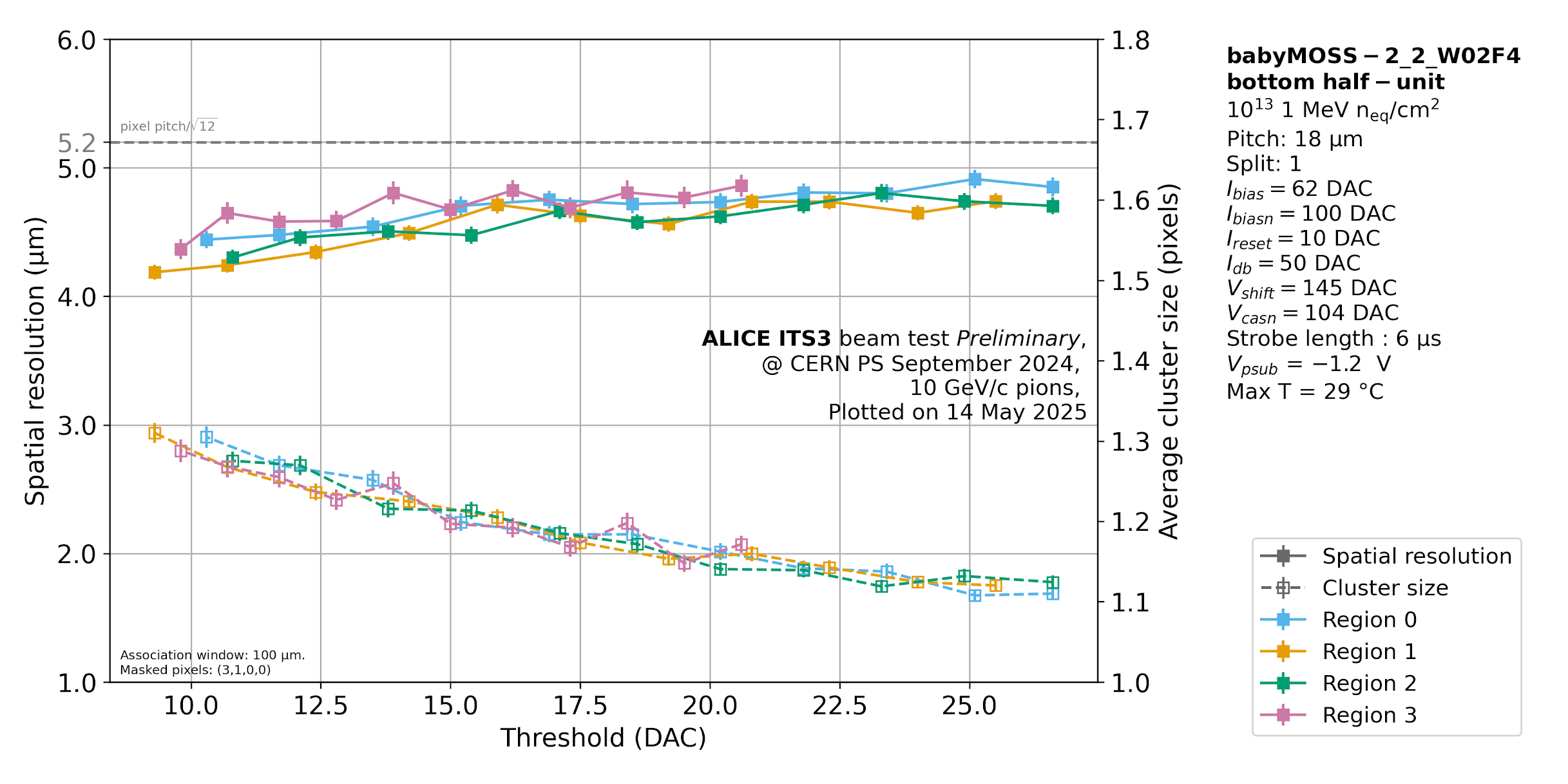}
	
	\caption{Spatial resolution (solid lines) and average cluster size (dashed) of all regions of the non-irradiated split~2 babyMOSS-2$\_$1$\_$W22C7 (top) and the irradiated split~1 babyMOSS-2$\_$2$\_$W02F4 (bottom), bottom half-unit.}
	\label{fig:res_noirr_vs_irr}
	
\end{figure}

Figure~\ref{fig:res_noirr_vs_irr} presents the spatial resolution and average cluster size of the non-irradiated and the irradiated devices, as functions of threshold. Both chips have performed well in terms of spatial resolution, falling below the binary limit in either device (5.2~$\upmu$m, dashed gray line).
However, in comparison with the non-irradiated chip, the irradiated DUT shows, to a limited extent, signs of spatial resolution degradation at very low thresholds, and smaller average cluster size. Indeed, non-ionising radiation can create trapping centres within the silicon lattice, affecting the charge collection process in our chip. The same effect was already observed in some small-scale test devices~\cite{rinella2023_dpts}, consistently with the findings presented here. At high thresholds, no appreciable differences in spatial resolution and cluster size were found. 

\section{Conclusions}
\label{sec:conclusions}
The babyMOSS chips are among the stitched test devices developed for the ALICE ITS3 upgrade studies, convenient for easy transportation and measurements that involve irradiation thanks to their small size ($\sim$14$\times$30~mm$^{2}$). BabyMOSS characterisation has been carried out in both laboratory and test beam environments: in particular, beam tests aim to evaluate the detection efficiency and FHR under a variety of different conditions, as well as to assess the radiation hardness of these devices. 
In September~2024, a test beam campaign was run at CERN PS, to assess the performances of 3~babyMOSS DUTs at varying irradiation levels, as 2 chips were pre-irradiated with neutrons at $10^{13}$~1~MeV~n$_{\mathrm{eq}}$cm$^{-2}$ NIEL level. This irradiation level did not seem to affect the detection efficiency, but has lead to increased FHR at thresholds above 20~DAC, due to higher leakage currents. Some degree of spatial resolution degradation and smaller cluster size were observed in the irradiated DUTs at low thresholds, but the performances remain comparable, and spatial resolution was found to be better than 5.2~$\upmu$m, regardless of the irradiation.

\end{document}